\documentclass[preprint]{siamart1116}

\usepackage{lipsum}
\usepackage{amsfonts}
\usepackage{graphicx}
\usepackage{epstopdf}
\ifpdf
\DeclareGraphicsExtensions{.eps,.pdf,.png,.jpg}
\else
\DeclareGraphicsExtensions{.eps}
\fi

\numberwithin{theorem}{section}

\newcommand{\TheTitle}{Benchmark problems for phase retrieval} 
\newcommand{\TheAuthors}{Veit~Elser, Ti-Yen~Lan and Tamir~Bendory}

\headers{\TheTitle}{\TheAuthors}

\title{{\TheTitle}}

\author{ Veit~Elser\footnotemark[1]\thanks{Department
		of Physics, Cornell University, Ithaca,
		NY, 14853-2501 USA
		(\email{ve10@cornell.edu}, \email{tl578@cornell.edu}).}
	\and
	Ti-Yen~Lan\footnotemark[1]
	\and
	Tamir~Bendory\footnotemark[2]\thanks{The Program in Applied and Computational Mathematics, Princeton University, Princeton, NJ, 08544-1000 USA
		(\email{tamir.bendory@princeton.edu}).}
}

\ifpdf
\hypersetup{
	pdftitle={\TheTitle},
	pdfauthor={\TheAuthors}
}
\fi

\usepackage[figuresright]{rotating}
\usepackage{multirow}
\usepackage{algorithm}
\usepackage{algpseudocode}
\usepackage{slashbox}
\usepackage{subcaption} 

\begin{document}
	
	\maketitle
	
\begin{abstract}
In recent years, the mathematical and algorithmic aspects of the phase retrieval problem have received considerable attention. 
Many papers in this area mention crystallography as a principal application. In crystallography, the signal to be recovered is periodic and comprised of atomic distributions arranged homogeneously in the unit cell of the crystal. The crystallographic problem is both the leading application and one of the hardest forms of phase retrieval. We have constructed a graded set of benchmark problems for evaluating algorithms that perform this type of phase retrieval. The data, publicly available online\footnote{\url{https://github.com/veitelser/phase-retrieval-benchmarks}}, is provided in an easily interpretable format.
We also propose a simple and unambiguous success/failure criterion based on the actual needs in crystallography.  
Baseline runtimes were obtained with an iterative algorithm that is similar but more transparent than those used in crystallography. Empirically,{the runtimes grow} exponentially with respect to a new hardness parameter: the sparsity of the signal autocorrelation. We also review the algorithms used by the leading software packages. This set of benchmark problems, we hope, will encourage the development of new algorithms for the phase retrieval problem in general, and crystallography in particular. \end{abstract}
	
	\begin{keywords}
phase retrieval, crystallography, periodic signals, reconstruction algorithms, benchmark problems, sparsity
	\end{keywords}
	
	\begin{AMS}
		
	\end{AMS}

\section{Introduction}\label{sec:intro}

The publication of ``Phase retrieval via matrix completion" by Cand{\`e}s et al.~\cite{candes2013phase} launched a revival of
interest in the phase retrieval problem. Phase retrieval originated in X-ray crystallography, which is still by far the largest application. In 2016,  
about 50,000 crystal structures were deposited in the Cambridge Structural Database~\cite{CSD}, each made possible by a phase retrieval algorithm. In brief, phase retrieval seeks to reconstruct a signal from measurements of its Fourier magnitudes. This is well-posed, of course, only if additional information is brought to bear on the reconstruction. In the case of X-ray crystallography, where the signal has support only at the positions of atoms, 
the additional information takes the form of a sparsity constraint.

All algorithms currently in use for crystallographic phase retrieval are close descendants of algorithms developed in the 1970s, 
and are viewed as heuristic methods by today's standards. By contrast --- see~\cite{bendory2017fourier,PR} for recent surveys --- the current wave of phase retrieval research has produced algorithms that can guarantee solutions. Is a migration to such algorithms
imminent in crystallography, or does crystallinity --- signal periodicity --- pose new, fundamental challenges?
In this paper, we make a case for the second scenario, and make contributions designed to help resolve what we see as an under-appreciated theoretical problem.

In Section~\ref{sec:crystallography}, we describe the crystallographic phase retrieval problem, highlighting those features that bring it outside the scope of current theoretical research and significantly increase its hardness. Our main contribution is presented in Section~\ref{sec:description}, where we describe a set of benchmark problems. These are synthetic instances of phase retrieval, designed with sufficient realism that success at solving them would translate to success with real data sets. Section~\ref{sec:construction} gives details on the benchmark constructions and, in particular, defines a new index, peculiar to crystallography, for ranking hardness. A representative heuristic method, described in Section~\ref{sec:baseline}, exhibits exponential time behavior with respect to this index. An easy way for a new brand of algorithm to distinguish itself is to demonstrate, at least empirically, behavior with smaller exponential growth or even sub-exponential behavior.

Heuristic methods in phase retrieval are often disparaged for lacking polynomial-time guarantees. This characterization, in the case of crystallographic phase retrieval, should be revisited if the new theoretical ideas fail to produce polynomial-time algorithms. As explained in Section~\ref{sec:baseline}, the (empirical) exponential behavior of the heuristic methods is still far superior to na\"ive exhaustive search. In addition, these methods appear to produce solutions reliably 
and their performance on the benchmarks in Section~\ref{sec:baseline} is strong motivation for their continued study.

Much of the current theoretical work on phase retrieval, for instance~\cite{candes2013phaselift,eldar2014phase,waldspurger2015phase}, was inspired by the proposal of Cand{\`e}s et al.~\cite{candes2013phase} of collecting a greater volume of magnitude-only data through the use of sensing vectors (designed or random) implemented as phase masks in the X-ray experiment.
This proposal overlooks a physical limit posed by the extreme smallness of the electron/X-ray interaction. For example, in a synchrotron X-ray diffraction experiment to determine the structure of a 30-nm virus particle, X-rays are elastically scattered into the detector (the source of all information) only at a rate of $10^{-1}$ X-rays per particle over the duration of the experiment.
Unless data is collected simultaneously from a very large number of identical particles, the number of bits of data arriving at the detector falls far short of what is required to define the particle's structure {(typically, the positions of many thousands of atoms)}. Therefore, data is always collected from crystals because it is the only feasible method of aligning the required large number of particles to enable the simultaneous recording from all of them. While periodicity {of the particles within the crystal} is clearly necessary for {an} adequate signal-to-noise ratio, the {exact same periodicity is required of the proposed phase mask} for obtaining additional magnitude-only data. In this setting, a ``phase mask" {would be a designed particle that co-crystallizes with the unknown target particle. This is a much used strategy for structure determination called ``molecular replacement," where the designed particle is a previously solved structure known to be a constituent part of the unknown particle. However, in molecular replacement the known information combines additively with the signal vector, not multiplicatively.} 
 
Because of physical limits that apply when resolving sub-nanometer detail, there is much less opportunity for designed measurement than at larger scales, such as is possible in medical imaging. Signal periodicity is therefore a non-negotiable feature of phase retrieval in structural biology. 
Section~\ref{sec:art} reviews four leading phase retrieval algorithms for biomolecular crystals. 
We do not present benchmark results on these algorithms but use this section to give an historical account of the evolution of phase retrieval algorithms, as exemplified by the method in Section~\ref{sec:baseline} for which we do provide detailed results. Section~\ref{sec:summary} summarizes the paper.

There are many non-crystallographic variants of the phase retrieval problem, arising in ptychography~\cite{thibault2008high,rodenburg2008ptychography}, laser pulse-shape characterization~\cite{trebino2012frequency}, etc., and we cannot hope to review them here. {However,} to satisfy the curiosity of the many researchers working on the simplified form of phase retrieval~\cite{candes2013phase}, we document in the appendix the behavior of our baseline algorithm (\textsf{RRR}) adapted for this aperiodic, no-prior-information case.

\section{Crystallography}\label{sec:crystallography}
A crystal is characterized by the periodic arrangement of a repeating structural unit, also known as the \emph{unit cell}. In X-ray crystallography, the ``signal" is the electron density function of the crystal,
\begin{equation}
\rho_c(\mathbf{x})=\sum_{\mathbf{y}\in S}\tilde{\rho}(\mathbf{x}-\mathbf{y}),
\end{equation}
where $\tilde{\rho}$ is a compactly supported motif and $S$ is
a finite set of translation vectors. The crystal parameters are defined by a lattice $\Lambda\subset\mathbb{R}^D$ and the set $S\subset \Lambda$ corresponds in practice to a very large and compact subset. If we fix a translation $\mathbf{z}\in\Lambda$ while increasing the size of $S$, the relation $\rho_c(\mathbf{x}+\mathbf{z})\approx\rho_c(\mathbf{x})$ becomes a better approximation, failing only for $\mathbf{x}$ near the surface of the crystal.

X-ray experiments measure the Fourier intensities $|\hat{\rho}_c(\mathbf{q})|^2$ of the crystal, where
\begin{align}
\hat{\rho}_c(\mathbf{q})&=\int_{\mathbb{R}^D}d\mathbf{x}~\rho_c(\mathbf{x})e^{-i 2\pi\mathbf{q}\cdot\mathbf{x}}\\
&=\left(\sum_{\mathbf{y}\in S}e^{-i 2\pi\mathbf{q}\cdot\mathbf{y}}\right)\left(\int_{\mathbb{R}^D}d\mathbf{x}'~\tilde{\rho}(\mathbf{x}')e^{-i 2\pi\mathbf{q}\cdot\mathbf{x}'}\right)\\
&=\hat{s}(\mathbf{q})~\hat{\rho}(\mathbf{q}).
\end{align}
As the set $S$ grows (includes more of $\Lambda$), the support of the function $\hat{s}(\mathbf{q})$ is increasingly concentrated on the dual lattice $\Lambda^*$. 

The $\mathbf{q}$-behavior of the measured Fourier intensities 
\begin{equation}
|\hat{\rho}_c(\mathbf{q})|^2=|\hat{s}(\mathbf{q})|^2~|\hat{\rho}(\mathbf{q})|^2
\end{equation}
is dominated by the function $|\hat{s}(\mathbf{q})|^2$: the phenomenon of \emph{Bragg peaks}. In this model of a perfect crystal,
the Bragg peak width is determined by the size of $S$. In practice, other factors (e.g., small misalignments of domains within the crystal) dominate the Bragg peak width. Nevertheless, the counterpart of $|\hat{s}(\mathbf{q})|^2$ for imperfect crystals will still have the periodicity of $\Lambda^*$ and comprise isolated peaks having widths on a scale so small that the function $|\hat{\rho}(\mathbf{q})|^2$ can be approximated as a constant over each peak. These features combine to give the integral of $|\hat{\rho}_c(\mathbf{q})|^2$, over just one Bragg peak, a very simple interpretation. First, since $|\hat{s}(\mathbf{q})|^2$ is periodic, its integral over a Bragg peak is independent of the specific peak and just contributes an overall scale factor. Second, the small width of the peaks ensures that the integrals are just sampling the function $|\hat{\rho}(\mathbf{q})|^2$ on the discrete set $\Lambda^*$. The X-ray measurements thus give the magnitudes of the Fourier series coefficients $\hat{\rho}(\mathbf{q})$, $\mathbf{q}\in\Lambda^*$, that define a strictly periodic function on $\mathbb{R}^D/\Lambda$:
\begin{equation}
\rho(\mathbf{x})=\frac{1}{\mathrm{vol}(\Lambda)}\sum_{\mathbf{q}\in\Lambda^*}\hat{\rho}(\mathbf{q})~e^{i 2\pi\mathbf{q}\cdot\mathbf{x}},
\end{equation}
where $\mathrm{vol}(\Lambda)$ denotes the volume of the crystal unit cell, and the phases of $\hat{\rho}(\mathbf{q})$ still have to be retrieved to reconstruct the density $\rho(\mathbf{x})$.

\begin{figure}[!t]
	\begin{center}
		\includegraphics[scale=0.6, trim=0cm 1cm 0cm 0cm, clip=true]{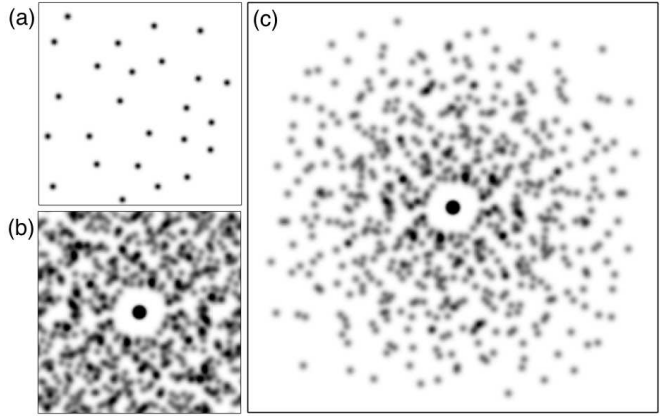}
	\end{center}
	\caption{(a) A signal $\rho$ comprising $N=25$ ``atoms." (b) The autocorrelation function of $\rho$ in the periodic case (the enclosing square is the unit cell). (c) The autocorrelation in the aperiodic case.}
	\label{fig:fig0}
\end{figure}

The phase retrieval problem can also be understood through autocorrelation analysis. In particular,
by the convolution theorem, the signal autocorrelation
\begin{equation}
a(\mathbf{y})=a(-\mathbf{y})=\int d\mathbf{x}~\rho(\mathbf{x})\rho(\mathbf{x}+\mathbf{y}),
\end{equation}
is given directly by the inverse Fourier transform of the signal's Fourier intensities $|\hat{\rho}(\mathbf{q})|^2$. It is instructive to compare the autocorrelation for an aperiodic signal, where $\rho$ and $a$ are defined on $\mathbb{R}^D$, to the periodic case where these are functions on $\mathbb{R}^D/\Lambda$. Figure~\ref{fig:fig0}(a) shows a signal comprising $N=25$ ``atoms" that may be interpreted either as the repeating motif of a crystal or the aperiodic signal of a single ``molecule." The autocorrelations $a(\mathbf{y})$ in both cases comprise $N(N-1)$ peaks associated with all pairs of atom-atom separations (and a strong peak at the origin for the $N$ self-correlations). In the periodic case, Figure~\ref{fig:fig0}(b), the same number of autocorrelation peaks is crowded into a smaller region than the aperiodic case, Figure~\ref{fig:fig0}(c). The resolution of the signal (atom size) and number of atoms in the illustration was set such that it is just becoming difficult to resolve individual autocorrelation peaks in the periodic case. By contrast, the aperiodic autocorrelation still has well resolved peaks, particularly at the periphery of the pattern, corresponding to the farthest separated atom pairs.

The structure of the autocorrelation $a(\mathbf{y})$, for a signal $\rho(\mathbf{x})$ comprising atoms, is helpful for understanding how periodicity increases phase retrieval hardness, that is, recovering $\rho(\mathbf{x})$ from $a(\mathbf{y})$. First, suppose that all the peaks in $a(\mathbf{y})$ are so well resolved that we know with high precision all the separation vectors, assumed to be unique. The atom positions in $\rho(\mathbf{x})$ can then be inferred by hand, starting with a pair of atoms having one of the unique separations, then placing atoms one at a time, each one constrained by having all separations to the placed atoms included among the unique separations in $a(\mathbf{y})$. This is a polynomial-time algorithm and applies equally in the periodic and aperiodic cases. But now consider the effects of finite resolution, that is, when the $O(N^2)$ number of separations outpaces the number of resolution elements, which normally scales as the number of atoms, $O(N)$. Errors arising from misidentified separations will occur with greater frequency in the periodic case because the autocorrelation peaks occupy a smaller region and have a higher density. Also, the density of peaks in the aperiodic case is not as uniform as it is in the periodic case, to the degree that sets of widely separated atoms can be reconstructed quite easily from the periphery of $a(\mathbf{y})$, where the peaks are sparse.

The structure of the autocorrelation function, now for the periodic case, also informs us on how phase retrieval hardness depends on parameters. Suppose we match the size of resolution elements to the size of atoms --- a reasonable approximation of what is done in crystallography. If our unit cell has $M$ resolution elements and $N$ atoms, we would characterize the signal sparsity as $N/M$. However, as a comparison of Figs.~\ref{fig:fig0}(a) and \ref{fig:fig0}(b) makes clear, a signal sparsity of $N/M$ 
translates to an autocorrelation sparsity of order $N^2/M$. 
Therefore, we should expect the hardness of the instances to scale like $N^2/M$ rather than the standard measure of signal sparsity.


An interesting consequence of the scaling difference between signal and autocorrelation sparsity is the possibility of instances that are both hard and have unique 
solutions. Many of the standard hard feasibility problems exhibit the phenomenon that the onset of solution non-uniqueness, as a function of a parameter in randomly generated instances, coincides with instances tuned to the maximum hardness. In some cases, e.g., logical satisfiability, the \textit{only} hard instances that can be generated randomly are right at the transition point. In crystallographic phase retrieval the situation is very different. We first note that the information deficit for a generic signal, defined as the ratio of the number of independent measurements to the number of independent signal values, is $1/2$. This follows from the fact that for a general real-valued signal $\rho(\mathbf{x})$ sampled on $M$ resolution elements there are $M/2$ independent Fourier magnitude samples $|\hat{\rho}(\mathbf{q})|$. By na\"ive constraint counting, a unique solution therefore requires at least $M/2$ real-valued constraints. But in the signals encountered in crystallography, Fig.~\ref{fig:fig0}(a), normally far more than half the samples are known, \textit{a priori}, to be zero.  
And, as we argued above, the retrieval of even such sparse signals pose a challenge when their autocorrelations are not sparse.

Phase retrieval has three fundamental, though benign, forms of solution non-uniqueness, frequently referred to as trivial ambiguities. The (real-valued) signal can be translated, inverted through the origin, or changed in sign, without changing the Fourier magnitudes. These ambiguities are associated with physical symmetries (arbitrariness of the sign of the electron charge, etc.) and do not compromise interpretability. A less trivial form of non-uniqueness, in the case of signals with well resolved atoms, arises when the geometry of the atom positions has the homometric property. Suppose $\rho$ and $\rho'$ are two solutions not related by one of the trivial ambiguities. Since they have equal Fourier magnitudes, they have matching autocorrelation functions and therefore share the same set of atom-atom separation vectors. In the aperiodic case, Rosenblatt and Seymour~\cite{rosenblatt1982structure} have shown that this only happens when $\rho$ is the convolution of (atomic) signals $\rho_1(\mathbf{x})$ and $\rho_2(\mathbf{x})$. When this is the case, a symmetry-unrelated signal $\rho'$ can be constructed from the convolution of $\rho_1(\mathbf{x})$ and $\rho_2(\mathbf{-x})$ that has the same autocorrelation. However, signals that ``factorize" in this sense are highly non-generic because there will be multiple symmetry-unrelated pairs of atoms that have exactly the same separation. In the generic case, when the peaks in the autocorrelation function all correspond to unique atom-atom separations, there will not be any homometric non-uniqueness. This conclusion also applies to the periodic case. A similar result was recently derived for the one-dimensional discrete case~\cite[Theorem 2.3]{beinert2015ambiguities}.

The relevance of periodicity for hardness can be argued on a more abstract level with the help of a discrete model, called \textit{bit retrieval}~\cite{E1}. In bit retrieval, a signal is represented by the integer coefficients of polynomials. The case where the non-zero coefficients are all equal to 1 comes closest to crystallography (in one dimension), where the polynomial $x^k$ corresponds to an atom located at position $k$. In the aperiodic case, the signal polynomials $b(x)$ are Laurent polynomials, $\mathbb{Z}[x,1/x]$, while in the periodic case (crystals) the polynomials belong to the ring $\mathbb{Z}[x]/(x^P-1)$ for some integer $P$ that defines the period. In bit retrieval we are given the autocorrelation polynomial $a(x)=b(x)b(1/x)$ and asked to find a factorization $a(x)=b'(x)b'(1/x)$, ideally such that $b'(x)$ also has all nonzero coefficients equal to 1. In the aperiodic case, the factors can be found in polynomial time by the LLL algorithm~\cite{lenstra1982factoring}, whereas no efficient factorization algorithms are known for the ring $\mathbb{Z}[x]/(x^P-1)$, the case of periodic signals.
 In fact, the hardness of factoring in the ring of cyclic polynomials is the basis of some cryptographic schemes~\cite{hoffstein1998ntru}. 

Crystals often have symmetries in addition to those conferred by the lattice of translations, $\Lambda$. Most generally, an (idealized) infinite crystal is unchanged by the action of the elements of a finite group $G$. 
An element $g\in G$ acts on the density function $\rho$ by the composition of a rotation $R_g$ (an orthogonal matrix) and a translation $T_g$:
\[
g\left(\rho(\mathbf{x})\right)=\rho(R_g\cdot\mathbf{x}+T_g).
\]
Thus in addition to
\[
\rho(\mathbf{x})=\rho(\mathbf{x}+\mathbf{y}), \quad \mathbf{y} \in \Lambda,
\]
the density of a crystal also satisfies
\[
\rho(\mathbf{x})=g\left(\rho(\mathbf{x})\right), \quad g\in G.
\]
The set of rotation matrices $R_g$ identify $G$ with a \textit{point group} (transformations that fix the origin), while the set of pairs $(R_g,T_g)$, together with the group of lattice translations $\Lambda$, specify the crystal's \textit{space group}. The order $|G|$ of the point group is usually denoted $Z$.

Space groups are usually identified prior to phase retrieval, often by the systematic vanishing of Fourier magnitudes $|\hat{\rho}(\mathbf{q})|$ for special $\mathbf{q}$. If $G$ is non-trivial, it has the effect of reducing the number of independent density samples in the crystal unit cell by the factor $Z$. Our benchmarks all have the trivial space group and avoid this complication.

\section{Description of the datasets}\label{sec:description}

In this section we describe our synthetic datasets and what it means to solve an instance. Details on the construction of the benchmark problems are given in the next section and can be skipped by readers just wishing to solve the benchmarks.

Data sets are identified by an integer $N$, the number of atoms in the signal, and a suffix characterizing the difficulty of the problem: E (easy), M (medium), H (hard). All data have the same format: an $M\times M$ table of integers  (photon counts) representing the measurements of Fourier intensities $|\hat{\rho}|^2$ of a signal $\rho$ sampled on a periodic $M\times M$ grid. All instances have $M=128$. This size was chosen to discourage methods~\cite{candes2013phase,huang2016phase} that represent the signal in terms of a dense, $M^2\times M^2$ matrix on which a rank-1 (or low rank) constraint is imposed. In 3-D protein crystallography the corresponding dense matrix could not be processed or even stored. Regarding the dimensionality of the signal, we believe this has no effect on phase retrieval complexity in the periodic case; two dimensions was chosen only for ease of rendering the signal. 

\begin{figure}[!t]
	\begin{center}
		\includegraphics[width=3.in]{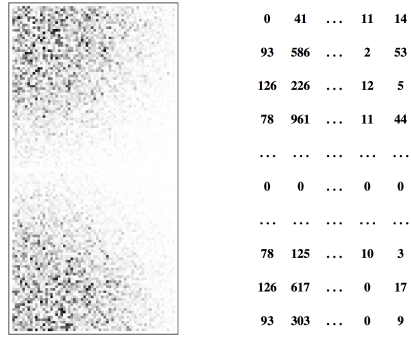}
	\end{center}
	\caption{Rendering (left) and excerpt (right) of benchmark instance \textbf{\texttt{data100E}}. The $(0,0)$ photon count at the upper left corner is not measured and set to zero.}
	\label{fig:fig1}
\end{figure}

Figure \ref{fig:fig1} shows a rendering and excerpt of the data file for the easiest instance, \textbf{\texttt{data100E}}. Though the data look random, there is a systematic decrease in the photon counts with increasing spatial frequency.
The table of photon counts contains several zero entries because, in experiments, intensities are normally measured out to frequencies where the Fourier transform is so small in magnitude that few photons are detected. The $(0,0)$ intensity (photons scattered with zero momentum change are indistinguishable from photons that did not scatter at all) is never measured and appears as a 0 in the data file. All other intensities have been symmetrized, $|\hat{\rho}(p,q)|^2=|\hat{\rho}(-p,-q)|^2$, because the electron density signal of X-ray crystallography is real-valued. The data files comprise just the  $128\times 64$ half-table of symmetrized photon counts. All entries in the $65^\mathrm{th}$ row and $65^\mathrm{th}$ column (not shown in Figure \ref{fig:fig1}) are zero.

Solving an instance entails the following. Square roots of the data are taken and define the Fourier magnitudes $|\hat{\rho}(p,q)|$. The phasing algorithm being demonstrated reconstructs $\hat{\rho}(0,0)>0$ and the phases $\phi(p,q)$ of the periodic signal
\begin{equation}\label{rho}
\rho(x,y)=\frac{1}{\sqrt{M^2}}\sum_{p=0}^{M-1} \sum_{q=0}^{M-1}e^{i\frac{2\pi}{M}(px+qy)}\;|\hat{\rho}(p,q)|e^{i\phi(p,q)}.
\end{equation}
The solution phases must satisfy $\phi(p,q)=-\phi(-p,-q)$ in order that $\rho(x,y)$ is real.

Solutions are required to be consistent with a prior constraint on the support $S$ of the signal. The cardinality of $S$ is exactly $8N$, that is, on average each of the $N$ atoms is supported on 8 pixels. In practice, we can take $S$ to be the set of pixels on which $\rho(x,y)$ has its $8N$ largest values because the signal is not only real but non-negative. Consistency with the support constraint is established by checking a power inequality. From the data, as well as the reconstructed $(0,0)$ intensity, the total Fourier power is given by
\begin{equation}\label{datapower}
I_\mathrm{F}=\sum_{p=0}^{M-1} \sum_{q=0}^{M-1}|\hat{\rho}(p,q)|^2 = \sum_{x=0}^{M-1} \sum_{y=0}^{M-1}{\rho}^2(x,y).
\end{equation}
In a successful reconstruction, the power in the support
\begin{equation}\label{supppower}
I_\mathrm{S}=\sum_{(x,y)\in S} \rho^2(x,y),
\end{equation}
matches the Fourier power. Because of Poisson noise (details in Section \ref{sec:construction}), the power in the support falls short of the power in the data. However, all the benchmark instances have solutions that satisfy:
\begin{equation}\label{solutioncriterion}
\frac{I_\mathrm{S}}{I_\mathrm{F}}>0.95.
\end{equation}

An instance is declared to be solved when this criterion is met. Understanding this criterion, and what makes it difficult to achieve, is the crux of the phase retrieval problem. Note that the denominator of \eqref{solutioncriterion} is mostly known. All that is unknown about $I_\mathrm{F}$ is the contribution of the constant term in the Fourier series, $|\hat{\rho}(0,0)|^2$. However, there is very little freedom in the value of this term when $\rho$ is required to be sparse. By contrast, the power in the support, $I_\mathrm{S}$, depends on all the unknown phases, via equations \eqref{supppower} and \eqref{rho}. Only very special combinations of phases, in combination with a special  $\hat{\rho}(0,0)$, can produce a $\rho$ that is highly sparse and for which the power in the small support (nearly) matches $I_\mathrm{F}$. Criterion \eqref{solutioncriterion} makes no reference to a ground-truth solution, nor does it make any assumptions about solution uniqueness.

Finally, we note that algorithms for solving the benchmarks are not required to work with signals sampled on $M\times M$ grids. Computations may be performed on finer or coarser grids, or without a grid sampling of the signal at all. However, at the end of the computation the algorithm is required to output the phase angles and $\hat{\rho}(0,0)$ needed to check criterion~\eqref{solutioncriterion}.

\section{Construction details}\label{sec:construction}

\begin{figure}[!t]
	\begin{center}
		\includegraphics[width=3.in]{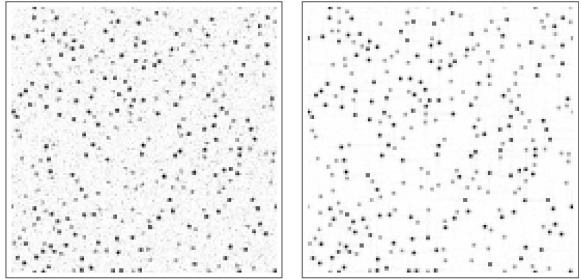}
	\end{center}
	\caption{A phase retrieval solution (left) and the as-constructed ground truth (right) for benchmark \textbf{\texttt{data300E}}. These images are rendered on a grid of the same size, $128\times 128$, as the grid that holds the data.}
	\label{fig:fig2}
\end{figure}

\begin{figure}[!t]
	\begin{center}
		\includegraphics[width=3.in]{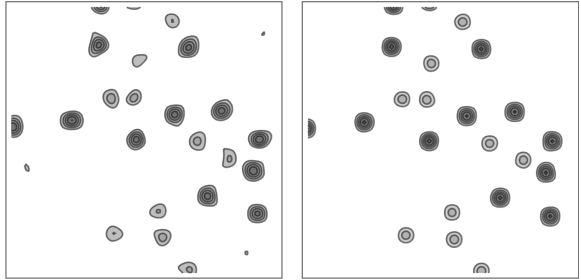}
	\end{center}
	\caption{Contour renderings of the upper left corners of the signals in Figure \ref{fig:fig2}.}
	\label{fig:fig3}
\end{figure}

Figure~\ref{fig:fig2} shows a successfully reconstructed signal for benchmark instance \textbf{\texttt{data300E}} next to the signal that 
was constructed, by the method described in this section, to produce the data (a translation and reflection through the origin was applied to the latter to aid comparison). Of the $128^2$ pixels, only $8\times 300$ have significant signal (appear gray). These are bandlimited signals and less noisy than they appear. {Smooth contour plots, formed by zero-padding the Fourier transforms on a $512\times  512$ grid and back transforming, are shown in Figure~\ref{fig:fig3}. We see ``atoms" of two types with centers having sub-grid precision.}

To construct data for an instance with $N$ atoms, first a set of $N$ atom-center pixels on a $512\times 512$ grid was sequentially sampled, uniformly but with the constraint that the distance between pixels is at least 12.
After downsampling the signal by a factor of $4$ {to avoid atoms artificially centered on grid points}, this gives a minimum separation of 3 pixels between atom centers, corresponding to 3 \AA{} in physical units (a typical atomic diameter) when the pixel resolution of the data is 1 \AA{} (typical of high quality data). 
The signal value was set to 1 on  $N/2$ {randomly selected pixels} and 2 on the other half. This mimics two species with atomic number ratio 2 and is designed to defeat algorithms that unrealistically impose an equal-atom prior. Minor variants of this part of the construction, described below, give the E, M and H grades of instances. 

After downsampling the signals from $512\times 512$ to $128\times 128$, the Fourier intensities (squared magnitudes) were multiplied by a Gaussian filter that had the effect of diminishing the lowest-frequency unmeasured intensities by a factor of 25 relative to intensities at the center of the transform. These filtered intensities were then rescaled (details below) and the result established the mean of the simulated photon counts, where the latter were sampled from the Poisson distribution. A single photon count was generated at each frequency, as in an actual experiment. The final step was to symmetrize the data by summing  the counts at $(p, q)$ and $(-p, -q)$.

The same intensity rescaling factor was used in all the benchmarks. This has the effect that individual atoms have the same characteristics (amplitude, width) across all the benchmarks. It also implies the total photon count in the data sets is proportional to $N$, the number of atoms. This normalization convention can be defended on information theoretic grounds: to reconstruct the types and positions (in a fixed field) of $N$ atoms, the quantity of information should be proportional to $N$. Since each detected photon delivers the same quantity of information, this proportionality is maintained when the total photon count is also proportional to $N$. 

Recall from section~\ref{sec:crystallography} that in crystallography the hardness of phase retrieval is expected to depend on the autocorrelation sparsity rather than the signal sparsity. Since the signal is comprised of $N$ atoms, the autocorrelation (away from the origin) is also ``atomic" in character, but with $N(N-1)$ peaks corresponding to all the interatomic vectors. These peaks are distributed homogeneously in the unit cell because the atoms themselves are distributed homogeneously. A small departure from the uniform distribution, see Figure~\ref{fig:fig0}(b), is explained by the constraint on the minimum atom-atom distance. For large $N$, a suitable definition of the autocorrelation sparsity, or the density of autocorrelation peaks, is therefore
\begin{equation}\label{mudef1}
\mu=\frac{N^2}{V},
\end{equation}
where $V$ is a measure of the number of pixels that can be resolved by the data.

The measure $V$ should correspond to the number of Fourier samples in the data, taking into account the decay in signal with increasing frequency. Because real data and our synthetic data is well characterized by Gaussian decay of intensity with frequency, we chose to define an ``effective" number of Fourier samples by the formula
\begin{equation}\label{volume}
V=\sum_{\mathbf{q}\in\Lambda^*}e^{-b|\mathbf{q}|^2},
\end{equation}
where the sum is over the  lattice dual to the crystal lattice $\Lambda$ and $b$ is a parameter. This definition applies in any number of dimensions. In real 3-D crystals, the intensity decay is reported as the value of the Wilson $B$-factor~\cite{G}, and $b=B/6$. For large $V$, the sum in \eqref{volume} can be approximated by an integral and we obtain, in three dimensions,
\begin{equation}
V\approx\left(\frac{6\pi}{B}\right)^{3/2}\mathrm{vol}(\Lambda),
\end{equation}
where the last factor is the volume of the crystal unit cell. 

Numerically performing the sum \eqref{volume} for the Gaussian filter of the benchmarks, we obtain the formula
\begin{equation}\label{mudef2}
\mu=\left(\frac{N}{64.17}\right)^2,
\end{equation}
for the benchmark instances. The numbers of atoms $N$ of the instances was chosen to sample $\mu$ by roughly equal intervals, from $\mu=2.4$ to $\mu=39$.

The benchmark signals all have trivial space group ($P1$), limiting comparisons with real-data phase retrieval. To expand the comparison group, we propose that in a space group with point group order $Z$, both $N$ and $V$ in \eqref{mudef1} should be divided by $Z$. This has the effect of replacing $\mu$ by $\mu/Z$. 
To the best of our knowledge and with this generalized definition, $\mu\approx 13$ is the hardest reported case of phase retrieval with real data for comparable structures (lacking heavy atoms; see Section \ref{sec:art}, Table~\ref{tab:tab2}).

When the atom positions are uniformly and independently sampled, the Fourier coefficients (before filtering) have (asymptotically) a complex-normal distribution by the central limit theorem. The corresponding intensities are then exponentially distributed, a phenomenon known as Wilson statistics~\cite{G}. This is a reasonable statistical model for the benchmarks, since the minimum distance constraints are rather weak for the densities of atoms considered. In real data it may happen that several intensities are unusually large by this model, and their existence can be exploited by clever algorithms. Conversely, phase retrieval appears to be more challenging when the data lacks such outliers. The extreme case was recently studied for two-valued one-dimensional signals~\cite{E1}, where it is possible to construct signals whose intensities are exactly equal. Since the intensity-distribution characteristics are clearly important, we implemented the following modification in the construction of the atom positions.

To quantify the outlier content of the intensity distribution, we used the normalized second-moment of the intensities 
\begin{equation}
i_2=\frac{\langle |\hat{\rho}|^4\rangle}{\langle |\hat{\rho}|^2\rangle^2},
\end{equation}
where $\langle{\cdot}\rangle$ denotes a uniform average over the measured, non-zero frequencies.
This statistical measure is increased when the high intensity tail of the distribution is enhanced, and decreases when the distribution is made more uniform.
Without any intervention, when atom positions are uniformly sampled (rejecting positions that violate the minimum distance constraint), we obtain $i_2\approx 4$. To make such instances easier, we select an atom at random and propose a new random position (still satisfying the distance constraint), accepting the proposal whenever the value of $i_2$ is increased. These increases are small, and many such moves had to be made to arrive at the value $i_2=4.5$ that define the E instances. The harder (H) instances were produced by the same procedure but where proposals are accepted whenever $i_2$ is decreased,  continuing until $i_2=3.5$. Relatively few atom-position re-samplings were needed to arrive at the value $i_2=4.0$ that defines our medium difficulty (M) instances.

\section{Baseline results}\label{sec:baseline}

To the best of our knowledge, the only known algorithms that reliably solve the benchmark problems are heuristic in nature. A common feature of these algorithms is that they act iteratively on the signal. To set a baseline for the benchmarks, we have selected a simple exemplar called Relaxed-Reflect-Reflect (\textsf{RRR})~\cite{E1}. This section describes the algorithm, addresses some common misconceptions about this type of algorithm, and suggests some standards for reporting results.

\algrenewcommand\algorithmicrequire{\textbf{input}}
\algrenewcommand\algorithmicensure{\textbf{output}}
\algrenewcommand{\algorithmiccomment}[1]{\hfill#1}

\begin{algorithm*}[t!]
	\caption{Relaxed-Reflect-Reflect (\textsf{RRR}) algorithm}
	\begin{algorithmic}[0]
		\Require $|\hat{\rho}|, |S|, \beta$\Comment{Fourier magnitudes, support size, \textsf{RRR} parameter}
		\State $\rho\leftarrow \mathrm{rand}()$ \Comment{random initial signal}
		\State $i\leftarrow 0$\Comment{zero the iteration counter}
		\Repeat
		\State $(\rho_1,S)\leftarrow P_1(\rho\;;|S|)$\Comment{support-size projection}
		\State $\rho_2\leftarrow P_2(2\rho_1-\rho\;;|\hat{\rho}|)$\Comment{Fourier magnitude projection}
		\State $\rho\leftarrow \rho+\beta(\rho_2-\rho_1)$\Comment{increment by the projection discrepancy}
		\State $i\leftarrow i+1$\Comment{increment counter}
		\Until{$\mathrm{pow}(\rho_2,S)>0.95$}\Comment{termination criterion~\eqref{solutioncriterion}}
		\Ensure $\rho_2, i$\Comment{phased input magnitudes (solution), iteration count}
	\end{algorithmic}\label{alg1}
\end{algorithm*}

Almost all crystallographic phase retrieval algorithms repeatedly use a ``Fourier synthesis" step, where a signal is constructed from the measured Fourier magnitudes and some estimate of the phases. The simplest such operation is the Fourier magnitude projection, $\rho\to\rho_2=P_2(\rho)$, where $\rho_2$ inherits its Fourier phases from an arbitrary signal $\rho$ and combines these with the Fourier magnitudes of the data, when available. When magnitude data is not available, say at frequency $\mathbf{q}$, the Fourier transform at $\mathbf{q}$ is simply copied.

Most algorithms also repeatedly do ``direct space refinement", where prior information is imposed on the signal. One of the simplest operations of this kind, when the signal is known to be sparse, is the support-size projection  $\rho\to\rho_1=P_1(\rho)$, where $\rho_1$ is unchanged on the $|S|$ highest valued pixels and set to zero on the rest --- in the process establishing the support $S$ of the signal. 
In \textsf{RRR}, the pixels belonging to the support $S$ are updated in each iteration.
We note that support-size projection is significantly weaker in constraining the signal than the analogous operation applied to a known support region. Going further, the case of a known support region $S$ that is sufficiently compact so it avoids aliasing (in the crystal unit cell) reverts to the easier, aperiodic phase retrieval problem. 

Pseudocode for~\textsf{RRR} is shown in Algorithm \ref{alg1}. In addition to the operations $P_2$ and $P_1$ already described, the algorithm calls on a function to initialize the signal and another function that computes the power ratio \eqref{solutioncriterion} for terminating iterations. Empirically, although the evolution of the signal is acutely sensitive to initial conditions (see below), the number of iterations required to find solutions, when averaged over different initial conditions, is not. Our implementation used a pseudo-random number generator for initialization. \textsf{RRR} has one parameter, $\beta$, that lies between 0 and 2.
Our baseline results are for $\beta=0.5$. 

From the formula for \textsf{RRR} iterations,
	\begin{equation}\label{eqn:RRR}
	\rho\mapsto\rho+\beta\left(P_2(2 P_1(\rho)-\rho)-P_1(\rho)\right),
	\end{equation}
we see that the parameter $\beta$ is like a time-step. However, the ``flow" in \textsf{RRR} does not correspond to a gradient, when $P_1$ and $P_2$ are interpreted as projectors to a general pair of constraint sets.
With $\beta=1$, the \textsf{RRR} algorithm~\cite{E1} generalizes, for arbitrary pairs of constraints, Fienup's ``hybrid input-output" (\textsf{HIO}) algorithm~\cite{F} for phase retrieval with a known support region, and coincides with the Douglas-Rachford splitting scheme when applied to partial differential equations~\cite{douglas1956numerical,bauschke2002phase}.

Many authors refer indiscriminately to \textsf{HIO} --- iteration \eqref{eqn:RRR} with $\beta=1$ --- and the algorithm that simply alternates the projections:
\begin{equation}\label{eqn:alternate}
\rho\mapsto P_2(P_1(\rho)).
\end{equation}
In the general convex setting this is ``von Neumann's alternating projections", while in microscopy and optics it is referred to as, respectively, ``Gerchberg-Saxton"~\cite{GS1972} and ``Fienup's error reduction"~\cite{F}. Aside from acting on the same space and being built from the same pair of projections, schemes \eqref{eqn:RRR} and \eqref{eqn:alternate} have almost nothing in common. Alternating projections is never used for hard (crystallographic) phase retrieval. In just a few iterations, it converges to one of a multitude of uninteresting fixed points: signals with correct Fourier magnitudes that are proximal to, but not coincident with, a signal of the correct support size. By contrast, when the \textsf{RRR} algorithm has a fixed point, it is because the correctly supported signal $\rho_1=P_1(\rho)$ is in the range of $P_2$ --- it also has the correct Fourier magnitudes.

\textsf{HIO}, as originally formulated~\cite{F}, also has a parameter $\beta$, but it does not correspond to a time-step and for $\beta\ne 1$, and non-linear $P_1$, loses the property that fixed points are automatically associated with solutions. We note that the support-size choice for $P_1$ in crystallography, unlike the known-support-region choice for \textsf{HIO}, is highly non-linear. Marchesini~\cite{marchesini2007phase} has proposed an optimization interpretation of \textsf{HIO} that applies for general $\beta$.
A predecessor of \textsf{RRR} was the ``difference map", a differently parameterized composition of the two projections with the property that reversing the sign of $\beta$ interchanges them~\cite{elser2003phase}. Constraint infeasibility, caused by noise in phase retrieval applications, was addressed by the ``relaxed averaged alternating reflection'' (\textsf{RAAR})~\cite{luke2004relaxed} modification of \textsf{HIO}.

It is also not accurate to characterize \textsf{HIO}/\textsf{RRR} as ``alternating" or ``cyclic" in the usual sense, as portrayed in popular accounts~\cite{PR}. To see this, note that a fixed point $\rho$ of \eqref{eqn:RRR} is in general not a solution. Instead it is the signals $\rho_1$ and $\rho_2$ generated by the projections and equal at a fixed point that are solutions (see \textsf{RRR} pseudocode)\footnote{If $\rho^*=\rho_1=\rho_2$ is a solution, then $\rho$ may be any point in the set $P_1^{-1}(\rho^*)\cap \left(2\rho^*-P_2^{-1}(\rho^*)\right)$.}. The truer sense in which this algorithm alternates is brought out by its equivalence, for $\beta=1$, to the ``alternating direction method of multipliers" (\textsf{ADMM}) algorithm~\cite{E2}.

Contemporary accounts often are dismissive of projection-based algorithms because convergence is not guaranteed, citing reports of ``stagnation" in the behavior of the iterates. While this criticism certainly applies to alternating projections, the direct opposite is empirically the case for \textsf{RRR}. The dynamics of \textsf{RRR} has the characteristics of a strongly mixing system in mechanics, where ergodic behavior is the rule rather than the exception. It is for this reason that initialization is unimportant. As in mechanics, the evidence of ergodicity is very strong even while prospects of a proof are dim. 
Ergodicity of the \textsf{RRR} dynamics would apply in a strict sense only for infeasible instances, when there are no fixed points. Because of noise, real-world instances are infeasible and the algorithm is remarkable in its ability to escape even near-solutions. Near-solution infeasibility is not an issue for finding phase retrieval solutions because we terminate the iterations as soon as the power ratio criterion~\eqref{solutioncriterion} is satisfied. Every one of our runs of \textsf{RRR} to solve a benchmark problem produced a solution: the success rate was 100\%.

\begin{figure}[!t]
	\begin{center}
		\includegraphics[width=3.in]{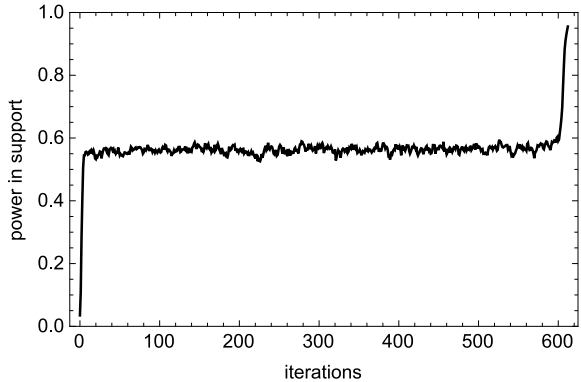}
	\end{center}
	\caption{Time series of the power ratio \eqref{solutioncriterion} over the course of solving \textbf{\texttt{data100H}} with \textsf{RRR}. Not only is the final capture by the solution fixed point very brief, so is the transient from the random initial signal to the family of signals explored in the search. Incorrectly phased signals in the long epoch of search have about 55\% power in a dynamic support constrained only by size.}
	\label{fig:fig4}
\end{figure}

There is no better illustration of the statistical behavior of the algorithm's search than the time series of the power ratio \eqref{solutioncriterion}. A typical time series for instance \textbf{\texttt{data100H}} is shown in Figure \ref{fig:fig4}. The solution --- marked by the sudden jump --- is not constructed incrementally but appears as an isolated event, when the chaotic dynamics arrives by chance at a fixed point's basin of attraction.
\textsf{RRR} is also used in convex optimization, where the behavior is very different and in fact convergent. However, when these algorithms are applied to hard phase retrieval only the final capture-phase of the solution process --- local convergence to the intersection of two affine sets --- falls under the purview of convex analysis.

\begin{figure}[!t]
	\begin{center}
		\includegraphics[width=3.in]{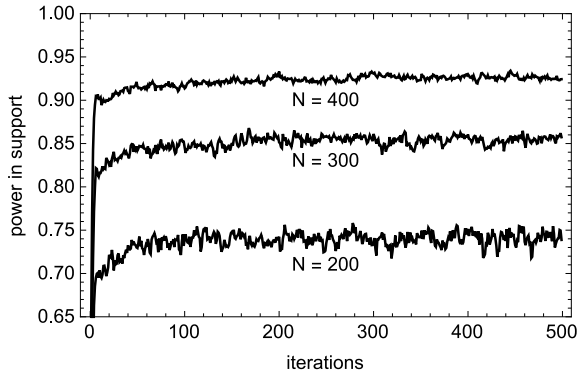}
	\end{center}
	\caption{Power ratio time series generated by \textsf{RRR} for instances with progressively more atoms showing just the steady state behavior prior to the solution-discovery jump seen in Figure \ref{fig:fig4}. Beyond 400 atoms the power-in-support fraction is too close to the value 0.95 (set by noise) to serve as a solution certificate.}
	\label{fig:fig5}
\end{figure}

The power ratio time series also illustrates the limits of using just the support size to constrain the signal. Figure \ref{fig:fig5} shows how the plot in Figure~\ref{fig:fig4} changes as the support size increases. The hardest benchmark instances, with $N=400$ atoms, are just short of the point where criterion \eqref{solutioncriterion} fails as a valid certificate. It is only for this reason that the benchmarks do not go beyond $N=400$ ($\mu=39$). Algorithms that seek signals with additional prior characteristics --- e.g., peaks --- could in principle succeed beyond this limit on the number of atoms. Indeed, algorithm developers are encouraged to exploit any of the prior signal information given in Section~\ref{sec:construction}. Criterion~\eqref{solutioncriterion} should only be seen as a convenient solution certificate that holds for $N\le 400$.

Benchmark results are most useful when behavior can be assessed with respect to hardness parameters. Though actual runtimes are important, trends in behavior are best reported in terms that do not depend on implementation details. In the case of \textsf{RRR}, the average number of iterations in repeated trials serves this purpose\footnote{The runtime per iteration in our implementation, about 1 msec, is essentially constant across all instances.}. Success rates, when greater than zero, are more useful when converted to an expected runtime per solution. Had \textsf{RRR} been run with a bound on the number of iterations, an expected runtime could have been computed from the total number of solutions found and the total number of iterations performed.

\begin{figure}[!t]
	\begin{center}
		\includegraphics[width=3.in]{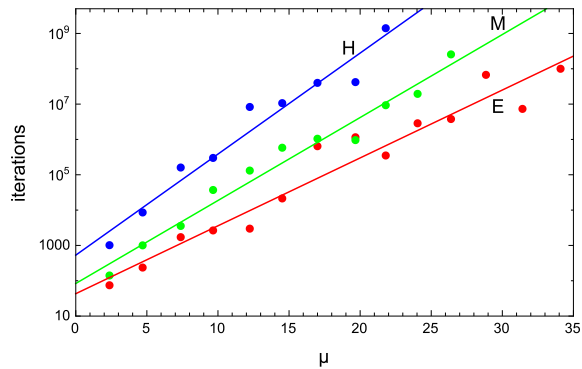}
	\end{center}
	\caption{Exponential growth of the mean iteration count for \textsf{RRR} as a function of $\mu$ for the three difficulty grades of instances. Linear fits to the logarithms of the iteration counts give the following factors by which the mean count grows when $\mu$ is increased by 1: 1.56 (E), 1.72 (M), 1.93 (H).}
	\label{fig:fig6}
\end{figure}

Iteration counts per solution for \textsf{RRR}, averaged over 20 trials per instance, are given in Table \ref{tab:tab1}. Figure \ref{fig:fig6} shows the behavior with respect to the autocorrelation sparsity parameter $\mu$ defined by \eqref{mudef2}. At each $N$, almost without exception, the E instance is easier than the M instance, which in turn is significantly easier than the H instance. The behavior with $\mu$ is consistent with a simple exponential. Linear fits to the logarithms of the iteration counts give the following factors by which the mean count grows when $\mu$ is increased by 1: 1.56 (E), 1.72 (M), 1.93 (H).

\begin{table}[!t]
	\renewcommand{\arraystretch}{1.3}
	\caption{\textsf{RRR} mean iteration counts (log$_{10}$)}
	\label{table:alg1esults}
	\centering
	\begin{tabular}{c|ccc}
		\hline
		$N$ & E & M & H\\
		\hline
		100 & 1.87 & 2.15 & 3.01\\
		140 & 2.37 & 3.00 & 3.93\\
		175 & 3.23 & 3.55 & 5.20\\
		200 & 3.42 & 4.57 & 5.48\\
		225 & 3.47 & 5.12 & 6.92\\
		245 & 4.33 & 5.77 & 7.03\\
		265 & 5.81 & 6.02 & 7.60\\
		285 & 6.06 & 5.98 & 7.62\\
		300 & 5.55 & 6.97 & 9.15\\
		315 & 6.46 & 7.29 & --\\
		330 & 6.58 & 8.41 & --\\
		345 & 7.83 & -- & --\\
		360 & 6.86 & -- & --\\
		375 & 8.00 & -- & --\\
		\hline
	\end{tabular}
	\label{tab:tab1}
\end{table}

We expect our baseline results to be an easy target. The best outcome of phase retrieval algorithm development would of course be a fundamentally different algorithm, promising subexponential cost in the parameter $\mu$. On the other hand, even incremental improvements in the case of exponential algorithms will bring substantial dividends. Although we have not performed the experiment, the extrapolation of our results indicate that \textsf{RRR} would require roughly three cpu-years to solve \textbf{\texttt{data330H}}, compared with the single hour needed for \textbf{\texttt{data330E}}.

\section{State of the art}\label{sec:art}

Probably the last time a phase retrieval milestone --- on real data --- was hailed was the Shake-and-Bake (\textsf{SnB}) solution of triclinic lysozyme in 1998 \cite{DM1}. The \textsf{SnB} algorithm was the product of a long history of developments that drew inspiration from various disciplines, including signal processing, probability theory and iterative methods for solving non-linear equations. In this section, we briefly review the principles behind \textsf{SnB}~\cite{SnB} as well as those used by three other leading crystallographic packages: \textsf{SHELXD}~\cite{SX}, \textsf{SIR2004}~\cite{SIR} and \textsf{SUPERFLIP}~\cite{SF}.

Because the earliest phase retrieval algorithms were developed in the pre-FFT era, they imposed prior information on the signal not directly in real-space, but indirectly during Fourier synthesis. The simplest such strategy, known as David Sayre's ``tangent formula"~\cite{TF}, is based on the observation that in a signal $\rho$ comprised of equal atom-like distributions (e.g., Gaussians), the Fourier phases of $\rho^2$ and $\rho$ are the same. This opens up the possibility that iterating $\rho\mapsto P_2(\rho^2)$ might by itself produce as fixed points a signal that (\textit{i}) has been synthesized from the known Fourier magnitudes and (\textit{ii}) corresponds to an atomic distribution --- at least for crystals of sufficiently identical atoms. It was possible to efficiently implement this map working just with the Fourier coefficients by expressing the transform of the square as the convolution of the transforms and then approximating the convolution by only the terms where both Fourier factors have a large magnitude. \textsf{SHELXD} and \textsf{SIR2004} have the option to apply the tangent formula operation in alternation with direct space refinement operations.

The tangent formula modification of the Fourier synthesis operation ($P_2$) is just one way the alternating scheme~\eqref{eqn:alternate} is made viable again, by eliminating a host of uninteresting fixed points. The identical-atom model of the signal on which the method is based is also the premise behind another modification of $P_2$. This is the \textsf{SnB} objective function on the phases that is first minimized before phases are combined with magnitudes~\cite{SnB}. The simplest form of the objective function is based on the observation that the distribution of the product $\hat{\rho}(\mathbf{q}_1)\hat{\rho}(\mathbf{q}_2)\hat{\rho}(\mathbf{q}_3)$ is invariant, for $\mathbf{q}_1+\mathbf{q}_2+\mathbf{q}_3=0$, under translation of all the atoms\footnote{This is also the underlying idea of analysis using the bispectrum~\cite{tukey1953,bendory2017bispectrum}.}. Therefore, it exhibits a non-trivial dependence on the sum of the corresponding phases, $\phi_1+\phi_2+\phi_3$. The distribution of this ``triplet" phase depends on the data via the known magnitude of the cubic product. The resulting conditional distribution for the triplet phase can be calculated explicitly for the case of equal atoms uniformly distributed in the crystal unit cell and serves as a model for triplet distributions in a typical crystal \cite{G}. \textsf{SnB} tries to bring the cosines of the triplet phases in line with their expectation in the random model by minimizing a sum-of-squares objective function.

It is important that modified Fourier synthesis, by either the tangent formula or the \textsf{SnB} objective function, is combined with a robust direct space refinement operation, such as the support-size projection $P_1$. This is because the phase interventions, or modifications of $P_2$, are based on approximate models and should only serve to bias the search for phases. When the bias built into $P_2$ has sufficient strength, one improves the probability that the signal $P_2(P_1(\rho))$ is also fixed by $P_1$ and is therefore a solution. After all, the bias in both methods (tangent formula, \textsf{SnB} objective function) is derived from a direct space model. Phase retrieval often succeeds simply by alternating a modified $P_2$ and a direct space refinement $P_1$ of some type. In \textsf{SnB}, \textsf{SHELXD} and \textsf{SIR2004} the $P_1$ operation can impose prior information on the signal beyond just the size of its support. This may include knowledge of the minimum atom-atom distance, the presence of a known number of heavy atoms, or even the expected histogram of the signal values.

The phase retrieval algorithm in \textsf{SUPERFLIP} also alternates between Fourier synthesis and direct space refinement, but unlike the other packages, it avoids false fixed points through a modification of the direct space operation~\cite{DM10}. When written in terms of the \textsf{RRR} projections, \textsf{SUPERFLIP} iterates a close approximation of the map
\begin{equation}
\rho\mapsto P_2(2 P_1(\rho)-\rho).
\end{equation}
The algorithm's name is derived from the argument of $P_2$, wherein the sign of the signal is reversed wherever it is judged not to be in the support and unchanged otherwise. In a solution, the ``charge flipping" step has no effect and the signal is a fixed point of the map. On the other hand, one cannot theoretically rule out the possibility of exotic non-solution fixed points, where charge flipping only changes the Fourier magnitudes --- that are then restored and the charge re-flipped by $P_2$. The $P_1$ used by \textsf{SUPERFLIP} \cite{DM10} differs from the one in \textsf{RRR} by being parameterized through a small positive lower bound on the signal value in the support rather than a bound on the support size. Although \textsf{RRR} is not based on an alternation of two operations, it is intriguing and perhaps not a complete coincidence that Fourier synthesis in this algorithm is also preceded by charge flipping.  

\begin{sidewaystable}
  \centering
  \begin{tabular}{lllclrlll}
		\hline
		\multirow{2}{*}{structure} & PDB or & \multirow{2}{*}{space group, $Z$} & \multirow{2}{*}{resolution (\AA)} & $N/Z$, & \multirow{2}{*}{$\mu/Z$} & iterations & \multirow{2}{*}{software} & \multirow{2}{*}{ref.}\\
			& CSD entry &	&	& heavy atoms &	& or time &		& \\
		\hline
		\hline
		hen egg white lysozyme & -- & $P1$, 1 & 0.85 & 1001, 10S & -- & 3,400 & \textsf{SnB} & \cite{DM1}\\
		alpha-1 peptide & 1BYZ & $P1$, 1 & 0.90 & 408, 1Cl & 1.73 & 4,500 & \textsf{SnB} &  \cite{DM2}\\
		acutohaemolysin & 1MC2 & $C2$, 4 & 0.85 & 975, 18S & 4.45 & -- & \textsf{SnB} &  \cite{DM3}\\
		scorpion toxin II & 1AHO & $P2_12_12_1$, 4 & 0.96 & 500, 10S & 4.46 & 413,000 & \textsf{SnB} &  \cite{DM4}\\
		\hline
		actinomycin D & 1A7Y & $P1$, 1 & 0.94 & 270 & 1.40 & -- & \textsf{SHELXD} &  \cite{DM5}\\
		feglymycin & 1W7Q & $P6_5$, 6 & 1.10 & 828 & 6.39 & -- & \textsf{SHELXD} &  \cite{DM6}\\
    		hen egg white lysozyme & 4LZT & $P1$, 1 & 0.95 & 1001, 10S & 11.06 & -- & \textsf{SHELXD} &  \cite{DM7}\\
		human cyclophilin G & 2WFI & $P2_12_12_1$, 4 & 0.75 & 1486, 2Mg 15S & 11.17 & 60 min &  \textsf{SHELXD} &  \cite{DM8}\\
		\hline
		hirustasin & 1BX7 & $P4_32_12$, 8 & 1.20 & 366, 12S & 4.29 & 546 min & \textsf{SIR2004} &  \cite{DM9}\\
		pheromone ER-1 & 2ERL & $C2$, 4 & 1.00 & 303, 8S & 4.72 & 19 min & \textsf{SIR2004} &  \cite{DM9}\\
		Kunitz domain C5 & 2KNT & $P2_1$, 2 & 1.20 & 460, 1P 6S & 6.85 & 22 min & \textsf{SIR2004} &  \cite{DM9}\\
		bovine ribonuclease & 1DY5 & $P2_1$, 2 & 0.87 & 1894, 31S & 12.53 & 131 min & \textsf{SIR2004} &  \cite{DM9}\\
		\hline
		2C$_{72}$N$_4$O$_6$ & PAWVEO & $P1$, 1 & 0.80 & 164 & 0.58 & 100 & \textsf{SUPERFLIP} &  \cite{DM10}\\
		2C$_{77.5}$N$_4$O$_{12.5}$ & GOFMOD & $P1$, 1 & 0.80 & 188 & 0.66 & 250 & \textsf{SUPERFLIP} &  \cite{DM10}\\
		apamin & -- & $P2_1$, 2 & 0.95 & 385 & 4.36 & 20,000 & \textsf{SUPERFLIP} &  \cite{DM11}\\
		\hline
	\end{tabular}

  \caption{Large, nearly equal-atom structures, solved by direct methods. PDB: Protein Data Bank. CSD: Cambridge Structural Database.}
  \label{tab:tab2}
\end{sidewaystable}

Direct comparison of the accomplishments of the crystallographic packages with the benchmark problems is complicated by a number of factors. First and foremost is the fact that data sets of sufficient quality, on which sparsity can be imposed, become increasingly rare as the number of atoms in the unit cell grows. In crystals of large protein molecules there is too much variability in the electron density from one unit cell to another (solvent disorder) that individual protein atoms cannot be resolved. The signal encoded by the Fourier magnitudes in this case is that of the average of the atomic distributions, and the corresponding broadening of the support has made it too weak to be used as a constraint. Almost all large protein structures are solved with the help of additional data, derived from atom-specific inelastic scattering. A large share of the credit for structures with $N$ above 1000 that did not rely on such additional data and yielded to ``direct methods" goes to the crystal growers who managed to significantly reduce disorder in their crystals.

Another factor that complicates comparisons is the presence of heavy atoms. Large proteins often contain a minority admixture of heavy atoms, and it is well known that this makes phase retrieval easier, even when this information is not used. This point is best explained by the structure of the autocorrelation. The presence of heavy atoms in a crystal produces strong autocorrelation peaks at the separations between the heavy atoms, from which the heavy atoms can first be located. Subsequently, the positions of the light atoms can be determined one at a time by the $O(N)$ number of autocorrelation peaks with intermediate signals, which correspond to the separations between a heavy atom and a light atom. The benchmark problems have equal numbers of atoms of two types and therefore correspond to harder instances in this respect.

Table \ref{tab:tab2} lists what we judge to be the hardest instances of successful real-data phase retrieval for crystals with (\textit{i}) resolved atoms and (\textit{ii}) the fewest number of heavy atoms. The highest value of $\mu$ in this list is near the middle of the benchmark problems.

\section{Summary}\label{sec:summary}

Phase retrieval can be decisive in the success of crystal structure discovery. Available algorithms for periodic signals are heuristic and their success and runtime behavior is poorly documented. It is not normal practice for crystallographers to test algorithms with synthetic data and a known ground truth; failures with real data are attributed to insufficient resolution or corrupted measurements, and usually go unreported. And when data quality is good, Nature does not always cooperate to create instances with graded hardness for the study of algorithm behavior. 

Phase retrieval theory moved into a new era of systematic study when it was taken up by applied mathematicians about seven years ago. However, the algorithms generated by this development have had no impact on phase retrieval for crystals. Periodicity of the signal in crystallography is not a minor property that can be recovered as a special case of the phase retrieval models that have been studied. 
In fact, the hardness conferred to the phase retrieval problem by periodicity seems to have completely escaped the notice of mathematicians.

Our benchmark problems were designed to address the circumstances just described. 
Crystallographers should evaluate their algorithms on standard benchmarks and applied mathematicians should turn their attention to phase retrieval problems that appear in real-world applications. The benchmark problems will serve both of these needs and, we hope, open a dialog between the two communities.

The recent creation of a software interface for phase retrieval, \textsf{PhasePack}~\cite{chandra2017phasepack}, illustrates the kind of problems that can arise when a healthy dialog is absent. One of the algorithms available in \textsf{PhasePack} is called ``Fienup", and a scientist working in the field would assume this is Fienup's \textsf{HIO} algorithm, the most widely adopted method and blueprint for the \textsf{RRR} generalization to general, non-convex, two-constraint problems. However, the ``Fienup" algorithm of \textsf{PhasePack} is just a minor variant of the alternating-projection (Gerchberg-Saxton) algorithm and only suitable for convex problems. It acquired this name, by accident it seems, because Fienup's article introducing \textsf{HIO} also featured the alternating-projection (``error-reduction") algorithm as a poor alternative.

Benchmarks are most useful when they uncover trends in behavior. We believe the autocorrelation sparsity parameter $\mu$, introduced here, will serve this purpose. The benchmarks instances sample this hardness parameter over a range that includes the frontier of real-world phase retrieval and the sampling is fine enough to be useful even when runtimes grow exponentially.

\section*{Acknowledgments}
V.E. and T-Y.L. received support from DOE grant DE-SC0005827. T-Y.L. was also supported by the Taiwan Government Scholarship to Study Abroad. T.B. would like to thank Amit Singer for his advice and support and Yonina Eldar and Mahdi Soltanolkotabi for helpful discussions.

\bibliographystyle{siamplain}
\bibliography{ref}

\appendix
\section{phase retrieval without prior information}

	Beginning with the publication of~\cite{candes2013phase}, ``phase retrieval" has become identified in the applied mathematics community with the following mathematical problem:
\begin{align}\label{eq:PR_simple}
		&\mbox{Given a sensing matrix }  A\in\mathbb{C}^{m\times n}\; \mbox{and magnitudes } |y|\in\mathbb{R}_{\ge 0}^m\;,\nonumber\\
		&\mbox{obtain a signal } \rho\in\mathbb{C}^n\; \mbox{such that } \vert y\vert = \vert A\rho\vert.
\end{align}
Under what conditions on $A$ can a signal that is unique, up to a global phase, be recovered, and how can this be done efficiently? This is conceptually simpler than the original, crystallographic, phase retrieval problem in that there are no prior constraints on $\rho$, such as non-negativity or sparsity. The loss of information on measuring magnitudes is entirely offset by having $m$ sufficiently large relative to $n$ for suitable~$A$.	

	Most of the current work on phase retrieval is directed at developing algorithms for solving \eqref{eq:PR_simple}. In this appendix we compare the performance of three leading algorithms with \textsf{RRR}, which is easily adapted to solve \eqref{eq:PR_simple}\footnote{A software interface for a wide range of phase retrieval algorithms is provided in~\cite{chandra2017phasepack}.}. 

The most direct real-world application of \eqref{eq:PR_simple} we are aware of is reconstructing the 2-D projected contrast of an isolated object from its diffraction intensity and knowledge of its 2-D support $S$. Suppose the measurements extend to spatial frequencies that resolve the contrast to pixels of area $r$. The signal vector $\rho$ will then have $n=|S|/r$ components. Consider the most common case, where these are real-valued. The diffracted intensity will then have centro-symmetry and, when sampled at resolution $r$, will be completely described by $m={\scriptstyle\frac{1}{2}}|S-S|/r$ measurements~\cite{elser2008reconstruction}. It is often possible to define $S$ by an enclosing mask of known size and shape. The success of algorithms depends critically on the ratio $m/n={\scriptstyle\frac{1}{2}}|S-S|/|S|$. For example, a triangular $S$ corresponds to $m/n=3$. 

The sensing matrix $A$ for the application just described would be an $m\times n$ submatrix of a 2-D Fourier matrix. The $n$ columns would correspond to the pixel positions in $S$, and the $m$ columns to detector measurement pixels, preferably optimized to improve the condition number of $A$. We are not aware of any demonstrations of~\eqref{eq:PR_simple} along these lines, for real or simulated data, by the algorithms recently proposed by the applied math community. This may simply be a reluctance to try the new methods on sensing matrices with ``structure." The comparisons in this appendix will therefore be with popular random models for $A$.

{In our first comparison, both the entries of $A$ and the ground truth $\rho_0$, in each instance,} were drawn from the normal distribution with zero mean and variance $1/2$ for both the real and imaginary parts. All algorithms had access to $A$ and the magnitudes $|y_0|=|A\rho_0|$. Since the signal $\rho$ can only be estimated up to a global phase, we define the normalized estimation error between $\rho$ and the ground truth $\rho_0$ by
	\begin{equation} \label{eq:error}
	\textrm{error} = \min_{\phi\in[0,2\pi)}\|\rho-e^{i\phi}\rho_0 \|_2/\|\rho_0\|_2.
	\end{equation}

	In the two-constraint formulation on which the \textsf{RRR} algorithm is based, we seek vectors $y\in\mathbb{C}^m$ which (\textit{i}) are in the range of $A$, and (\textit{ii}) have the known magnitudes $|y_0|$. The corresponding constraint projections are
	\begin{align}
	P_1(y)&=A A^\dag\, y,\\
	P_2(y)&=|y_0| e^{i\arg{y}},
	\end{align}
where $A^\dag=(A^* A)^{-1}A^*$ is the pseudo-inverse of $A$. The \textsf{RRR} parameter $\beta$ was set at 0.5 and $\rho$ was initialized naively, each element drawn from the complex normal distribution. Iterations were terminated when the normalized norm of the difference of successive iterates, $\|y'-y\|_2 / \|y\|_2$, dropped below $10^{-8}$. 
The resulting $y$ is approximately fixed by both projections and gives the solution estimate $\rho=A^\dag\, y$. Our solver was written in \textsf{Matlab} and did not require any special packages.

	We compared \textsf{RRR} with three recently proposed algorithms designed to solve instances of~\eqref{eq:PR_simple}. 
	The first method, called \textsf{PhaseLift}, is based on the insight that~\eqref{eq:PR_simple} can be formulated as a set of linear equations with respect to the rank-1 Hermitian matrix $\rho\rho^*$. Applying a standard convex relaxation to the rank-1 constraint leads to a semidefinite program (SDP) that can be solved in polynomial time. The signal is estimated as the leading eigenvector of the SDP's solution. This technique has been thoroughly analyzed in a series of papers; see for instance~\cite{candes2013phase,candes2013phaselift,candes2014solving,candes2015phasecode,waldspurger2015phase}. We used \textsf{CVX toolbox}~\cite{grant2008cvx} to solve the SDP.
	
	The second algorithm, Truncated Wirtinger Flow (\textsf{TWF}), was proposed in~\cite{chen2017solving} and minimizes a non-convex least-squares objective by a gradient algorithm. The initial signal estimate is generated by a spectral algorithm. We used the implementation provided by the authors\footnote{\url{http://statweb.stanford.edu/~candes/TWF/code.html}} with $100$ iterations for the initialization and $2000$ gradient iterations, with step size $0.2$, for minimization. Several papers have proposed similar techniques based on different objective functions; see for instance~\cite{wang2017solving,zhang2017nonconvex}. However, we found these modifications had little effect on performance.   

	The last algorithm, proposed in~\cite{bendory2017non}, is called Non-Convex PhaseCut (\textsf{NCPC}) and is based on the PhaseCut algorithm~\cite{waldspurger2015phase}. Whereas the original PhaseCut algorithm used the classical Max-Cut SDP relaxation~\cite{goemans1995improved}, the \textsf{NCPC} algorithm minimizes instead a non-convex objective on the manifold of phases. In particular, \textsf{NCPC} employs a Riemannian trust-regions algorithm using the \textsf{Manopt toolbox}~\cite{boumal2014manopt}. The signal is initialized by the spectral algorithm of~\cite{chen2017solving} with $100$ iterations. We interrupted the trust-regions iterations when the norm of the gradient dropped below $10^{-6}$.

\begin{figure}
	\centering
	\includegraphics[scale=.6]{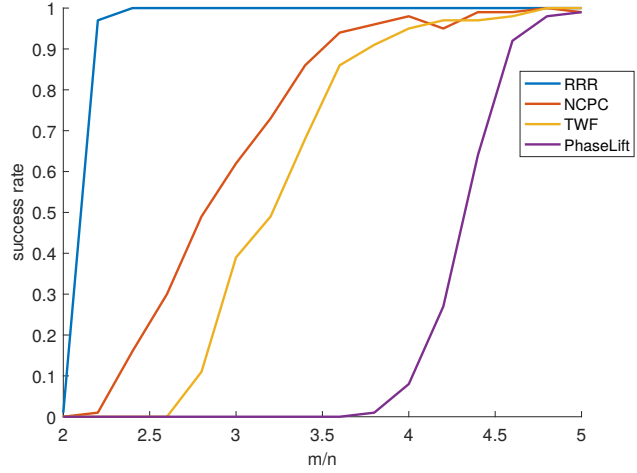}
	\caption{The success rates of the \textsf{RRR}, \textsf{NCPC}~\cite{bendory2017non}, \textsf{TWF}~\cite{chen2017solving}, and \textsf{PhaseLift}~\cite{candes2013phase} algorithms as a function of $m/n$ for problem~\eqref{eq:PR_simple} with $n=50$. A trial was declared successful if the error~\eqref{eq:error} was below $10^{-7}$.}
	\label{fig:success_rate}
\end{figure}

	Figure~\ref{fig:success_rate} plots the success rates of the four algorithms when recovering complex signals of length $n=50$ as a function of the ratio $m/n$. For each $m/n$, 100 trials were performed and a trial was declared successful if the error~\eqref{eq:error} was below $10^{-7}$. The same $\rho_0$ and $A$ were given to all the algorithms in each trial. Recall that there are $2n$ parameters in a successful reconstruction, the real and imaginary parts of the signal. As can be seen, \textsf{RRR} is the only algorithm that achieves a perfect success rate close to this information limit. 

	Table~\ref{tab:wall_clock_per_recovery} gives the total time used by each algorithm over all 100 trials, divided by the number of successful reconstructions. These values should be interpreted as the average time needed to achieve a successful reconstruction. We see that \textsf{RRR} outperforms the other algorithms over all the $m/n$ values considered. These results should be taken with caution since the algorithms used different stopping criteria. For instance, the \textsf{TWF} code we adopted from the authors' website halts after a given number of iterations (in our case, $2000$). By contrast, the stopping criteria of the other algorithms are based on the progress of the algorithm (e.g., when the  norm of the gradient is smaller than some predefined value.) The parameters for the stopping criteria (reported above) were not optimized to minimize the time per successful recovery.

\begin{table}
	\begin{center}
		\begin{tabular}{ |l||r|r|r|r| } 
			\hline
			 \backslashbox{algorithm}{$m/n$}& 3 & 4 &  5  \\ \hline \hline 
			\textsf{RRR} & 0.155 & 0.169 & 0.230  \\ \hline
			\textsf{NCPC} & 0.405 & 0.252  & 0.269  \\ \hline
			\textsf{TWF} & 0.780 & 0.352 & 0.357  \\ \hline
			\textsf{PhaseLift} & ---  & 186.4 & 19.4  \\ \hline		
		\end{tabular}
		\caption{Average CPU time in seconds per successful reconstruction for different $m/n$ values. 
		\label{tab:wall_clock_per_recovery}}
	\end{center}
\end{table}

	Cand{\`e}s et al.~\cite{candes2013phase} also studied instances of problem~\eqref{eq:PR_simple} where the sensing matrix $A$ corresponds to the following case of the coded diffraction pattern (CDP) model. Given $L$ random phase masks whose entries $d_\ell(s, t)$ are drawn independently and uniformly from $\{-1, 1, i, -i\}$, the resulting measurements are
	\begin{equation} \label{eq:cdp}
	|y_\ell(p, q)| = \bigg| \sum_{s=0}^{M-1}\sum_{t=0}^{N-1}\bar{d}_\ell(s, t) \rho(s, t) e^{-2\pi i ps/M} e^{-2\pi i qt/N} \bigg|,\quad \ell = 1, \dots, L,
	\end{equation}
where $\rho$ is the $M \times N$ signal to be reconstructed. The integer $L$ determines the redundancy of the measurements, corresponding to the ratio $m/n$ in the paragraphs above. By combining indices, $\{\ell,p, q\}\to\mu$ and $\{s, t\}\to\nu$, the entries of the sensing matrix $A$ are given by
	\begin{equation} \label{eq:sensing}
	A_{\mu\nu} = \bar{d}_\ell(s, t) e^{-2\pi i ps/M} e^{-2\pi i qt/N}.
	\end{equation}

In order to test the \textsf{RRR} algorithm on CDP problems, we have to efficiently implement the projection $P_1(y)=A A^\dag\, y$. This is done by exploiting the fast Fourier transform (FFT). From equation~(\ref{eq:sensing}), we can readily obtain the relations
	\begin{align}
	(A\rho)_\mu &= \mathcal{F}\big\{\bar{d}_\ell(s, t) \rho(s, t) \big\}(p, q) \\
	(A^* y)_\nu &= \sum_{l=1}^L\sum_{p=0}^{M-1}\sum_{q=0}^{N-1} d_\ell(s, t) e^{2\pi i ps/M} e^{2\pi i qt/N} y_\ell(p, q) \nonumber \\
		&= MN~\sum_{l=1}^L d_\ell(s, t)~\mathcal{F}^{-1}\big\{y_\ell(p, q) \big\}(s, t),
	\end{align}
where $\mathcal{F}$ and $\mathcal{F}^{-1}$ denote the 2-D FFT and its inverse\footnote{Here we use the asymmetric normalization convention for the discrete Fourier transform to be consistent with the FFT codes in \textsf{Matlab}. The unitary convention was adopted in the main text.}. Moreover, the product $A^*A$ is a diagonal matrix with diagonal terms
	\begin{equation}
	(A^*A)_{\nu\nu} = MN\sum_{l=1}^L |d_\ell(s, t)|^2.
	\end{equation}
The projection $P_1(y)=A A^\dag\, y = A(A^*A)^{-1}A^*\, y$ can therefore be implemented via pairs of FFTs, without explicit construction of the matrices. 

\begin{figure}[t]
\centering
\includegraphics[scale=0.72, trim=9cm 2.5cm 10cm 3cm, clip=true]{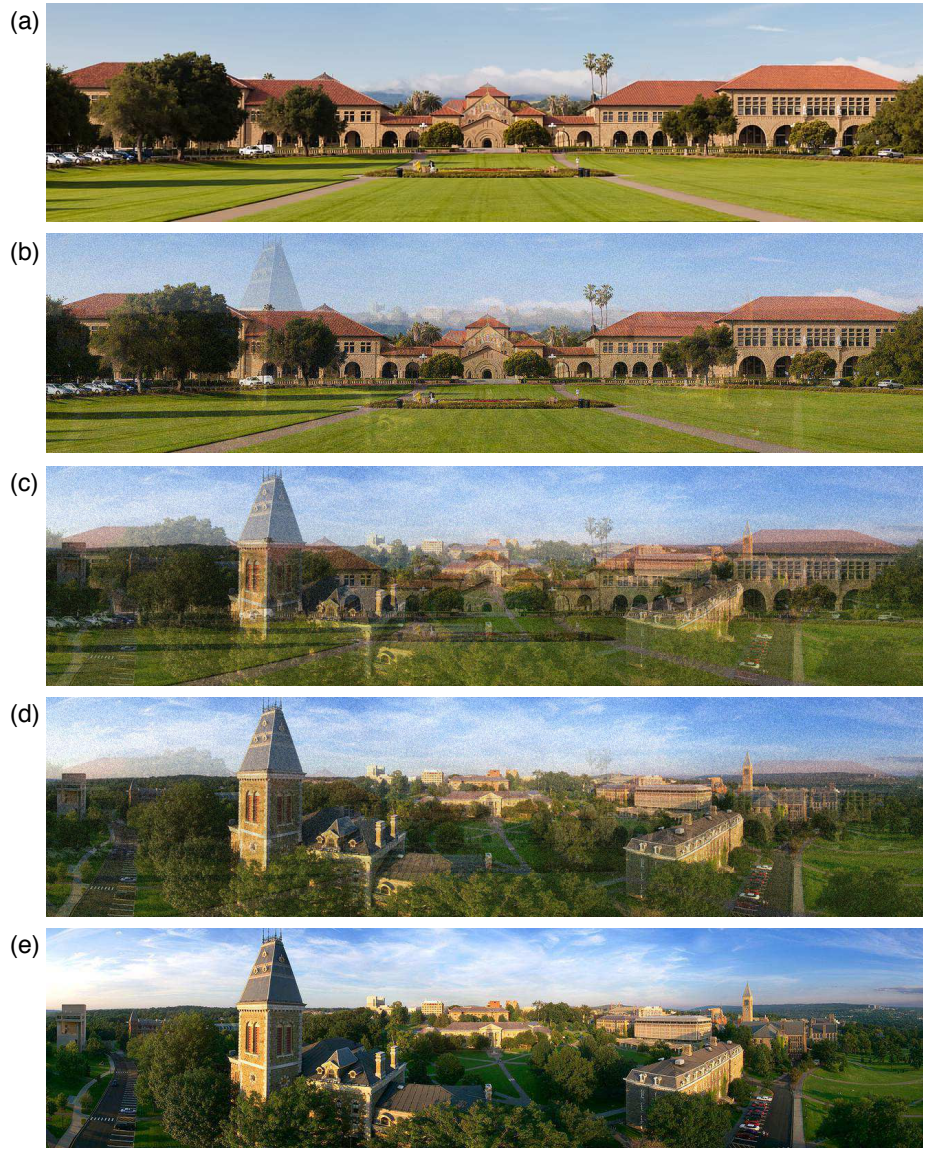}
\caption{Reconstruction of the Cornell University image using the \textsf{RRR} algorithm on the CDP model~\eqref{eq:cdp} with $L=3$. The top panel (a) is the initial guess (Stanford University) and the following panels (top to bottom) show the convergence of the reconstruction after (b) 1 (c) 3 (d) 5 (e) 250 \textsf{RRR} iterations.}
\label{fig:Cornell_recovery}
\end{figure}

\begin{figure}
	\centering
	\includegraphics[scale=0.45]{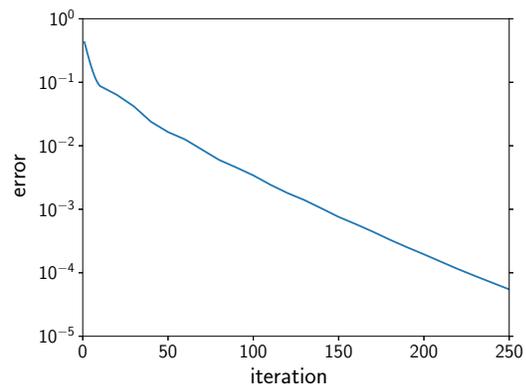}
	\caption{Reconstruction error~\eqref{eq:error} vs. iterations for the phase retrieval experiment shown in Figure \ref{fig:Cornell_recovery}.}
	\label{fig:Cornell_error}
\end{figure}

Following tradition, we demonstrate CDP phase retrieval on photographic images\footnote{This is not the best practice, because the Fourier power in photographs is artificially concentrated along the lines $p=0$ and $q=0$ due to image termination (aperiodicity).}. 
Our test of \textsf{RRR} used only $L=3$ masks, where previous work~\cite{chen2017solving} used $L=12$. For the ground truth signal we used three $320 \times 1280$ images of Cornell University, one for each color band (red, green and blue). To confuse the algorithm we used an image of Stanford University as the initial signal. Figure~\ref{fig:Cornell_recovery} (b) - (e) show the \textsf{RRR} image $A^\dag\, P_1(y)$ after $1,3,5$ and $250$ iterations. The error~\eqref{eq:error} after 250 \textsf{RRR} iterations is $5.48\times10^{-5}$. As can be seen, \textsf{RRR} needs only a few iterations to get a fine estimation of the signal. Figure~\ref{fig:Cornell_error} gives the reconstruction error~\eqref{eq:error} as a function of the iterations. CDP reconstructions with as few as a single random phase mask, using \textsf{HIO} or the Douglas-Rachford algorithm, have been demonstrated when prior constraints on the signal are also used~\cite{fannjiang2012phase,chen2016fourier}.

\end{document}